# Analyses of femtosecond laser ablation of Ti, Zr and Hf.

D. Grojo, J. Hermann[*], S. Bruneau and T. Itina

LP3-FRE 2165 CNRS, Faculté des Sciences de Luminy, Case 917, 13288 Marseille Cedex 9, France

**ABSTRACT**

Femtosecond laser ablation of Ti, Zr and Hf has been investigated by means of in-situ plasma diagnostics. Fast plasma imaging with the aid of an intensified charged coupled device (ICCD) camera was used to characterise the plasma plume expansion on a nanosecond time scale. Time- and space-resolved optical emission spectroscopy was employed to perform time-of-flight measurements of ions and neutral atoms. It is shown that two plasma components with different expansion velocities are generated by the ultra-short laser ablation process. The expansion behaviour of these two components has been analysed as a function of laser fluence and target material. The results are discussed in terms of mechanisms responsible for ultra-short laser ablation.

**Keywords:** Laser ablation, femtosecond laser, plasma spectroscopy

## 1. INTRODUCTION

Ultra-fast laser ablation is of growing interest pushed by applications like high precision micromachining, micro-analysis of materials, pulsed laser deposition of thin films, etc … . Compared to ablation with nanosecond laser sources, the thermal damage of the material is strongly reduced when using ultra-short laser pulses because of the much lower thermal heat diffusion into the material bulk [1]. Several theoretical models were proposed to describe ultra-fast laser ablation. These models involve different ablation mechanisms like thermal ablation [2], electrostatic ablation [3], Coulomb explosion [4], phase explosion [5], etc … . Although significant progress has been obtained during the past few years in the research field of ultra-fast laser-matter interaction, there is still a lack of knowledge. In this context, the present study is an experimental approach, based on analyses of the laser-induced plasma, to get a better insight in the fundamental physics of ultra-fast laser ablation.

## 2. EXPERIMENT

A scheme of the experimental arrangement is given in figure 1. Ablation experiments have been performed using a Ti:Sapphire laser (Spectraphysics, model Hurricane) delivering pulses of 100 fs duration, 1 mJ energy at a repetition rate of 1 kHz. A spot of uniform energy distribution and 30 µm diameter has been obtained on the target surface by imaging a 1.5 mm aperture by a lens of 50 mm focal length. The laser beam energy was varied with the aid of calibrated attenuating plates. The number of applied laser pulses has been controlled with the aid of a mechanical shutter. Both the fundamental laser wavelength ($\lambda_{las}$ = 800 nm) and the second harmonic ($\lambda_{las}$ = 400 nm) have been used in the experiments. The second harmonic was generated by a B.B.O. crystal. The metal targets were placed in a stainless steel vacuum chamber with a $10^{-4}$ Pa residual pressure. The vacuum chamber is equipped with 4 quartz windows for the laser beam pathing through and optical accessing. Inside the chamber, target holder and focusing lens are mounted on motorised translation axes. The focusing distance and the target position were controlled with the aid of a CCD camera by capturing the target surface through the focusing lens. The latter was protected by thin glass plates against deposition of ablated material. The glass plates were changed regularly to avoid perturbation of the laser beam by the deposit. During most of the experiments, the chamber was filled with He at 500 Pa pressure. The background gas was introduced in order to reduce the deposition rate on the glass plates. However, the pressure was low enough to avoid perturbation of the ablation process by the He atmosphere.

Fast plasma imaging has been performed with the aid of a focusing objective (Tamron 70-210 mm, 1:3.8-4) and a fast intensified charge coupled device (ICCD) (Andor, iStar). The temporal and spatial resolution of these measurements were about 5 ns and 60 µm, respectively. For spectroscopic observations, the plasma plume was imaged on the entrance slit of a 1 m spectrometer (Jobin-Yvon). Using two achromatic quartz lenses of 200 and 400 mm focal length, an image magnification ×2 was obtained. According to the width of 100 µm of the spectrometer entrance slit, the plasma emission was captured from a zone of a dimension of about 50 µm in the direction parallel to the laser beam and plasma symmetry axis. The distance $z$ of the observation zone with respect to the target surface varied by displacing the $f = 400$ mm lens along the $z$-axis. A fast intensified charge coupled device (Princeton Instruments, model 576/RB-E) with a variable gate (≥ 5 ns) was placed at the spectrometer output for photon detection. Using the first order of the spectrometer grating, a spectral window of about 6 nm was observed with a spectral resolution of $\lambda/\Delta\lambda = 1\times10^4$ ($\Delta\lambda = 0.05$ nm at $\lambda = 500$ nm). The spatial resolution of the lateral $x$–position, perpendicular to the laser beam axis, was about 22 µm. Synchronisation between laser and ICCD detector was ensured using the Pockels cells pulse generator trigger output of the femtosecond amplifier. The time jitter between laser and detector gate was about ± 1 ns. ICCD detector and spectrometer were connected to a microcomputer for data acquisition and spectra analysis. In order to enhance the signal to noise ratio, data acquisition was performed by averaging the signal over $10^2$ or more successive laser shots. During the recordings, the target was translated perpendicularly to the laser beam with a velocity of 30 µm s$^{-1}$ in order to maintain the laser-material interaction stable. An observation gate of 5 ns or 10 ns was used during most experiments. The time delay $t$ between laser pulse and observation gate was varied with the aid of a delayed pulse generator. According to the available maximum energy of 20 µJ at 800 nm, a maximum fluence of 3 Jcm$^{-2}$ was obtained. For the second harmonic, the maximum fluence was 1.6 Jcm$^{-2}$.

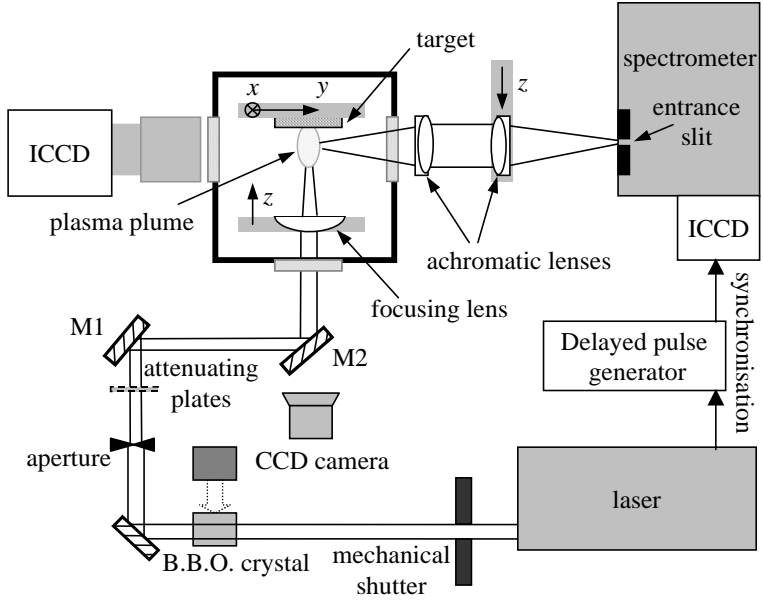

Figure 1 : Experimental set-up. Mirrors M1 and M2 with coatings for reflection at 45° incidence at $\lambda_{las}$ (800 or 400 nm) were used.

## 3. RESULTS AND DISCUSSION

### 3.1. Fast plasma imaging

The thermo-physical properties of Ti, Zr and Hf are listed in table 1. It is noted that the three metals, which are situated in the same column of the periodic table of elements, have similar optical and thermo-physical properties, but different atomic mass. Most of the experiments have been carried out in He background gas of 500 Pa pressure for two reasons. First, the presence of the background gas leads to a reduced deposition rate on the lens protecting glass plates. Second, collisions between the ablated vapour species and atoms lead to higher excitation rates and therefore a higher detection efficiency of the plasma species. However, it is emphasised that the pressure was low enough to avoid perturbation of the ablation process by the He atmosphere.

| Species | $M$ (a.m.u.) | $\rho$ (g cm$^{-3}$) | $k_{th}$ (W m$^{-1}$ K$^{-1}$) | $C_P$ (J K$^{-1}$ kg$^{-1}$) | $T_{vap}$ (K) | $T_{fus}$ (K) | $L_{vap}$ (J g$^{-1}$) | $L_{fus}$ (J g$^{-1}$) |
|---|---|---|---|---|---|---|---|---|
| Ti | 47.9 | 4.5 | 21.9 | 523 | 3560.15 | 1933.15 | 8893 | 365 |
| Zr | 91.2 | 6.49 | 22.7 | 281 | 4650.15 | 2125.15 | 6360 | 211 |
| Hf | 178.5 | 13.1 | 23 | 146 | 4875.15 | 2500.15 | 3700 | 122 |

Table 1 : Thermo-physical properties of Ti, Zr, Hf

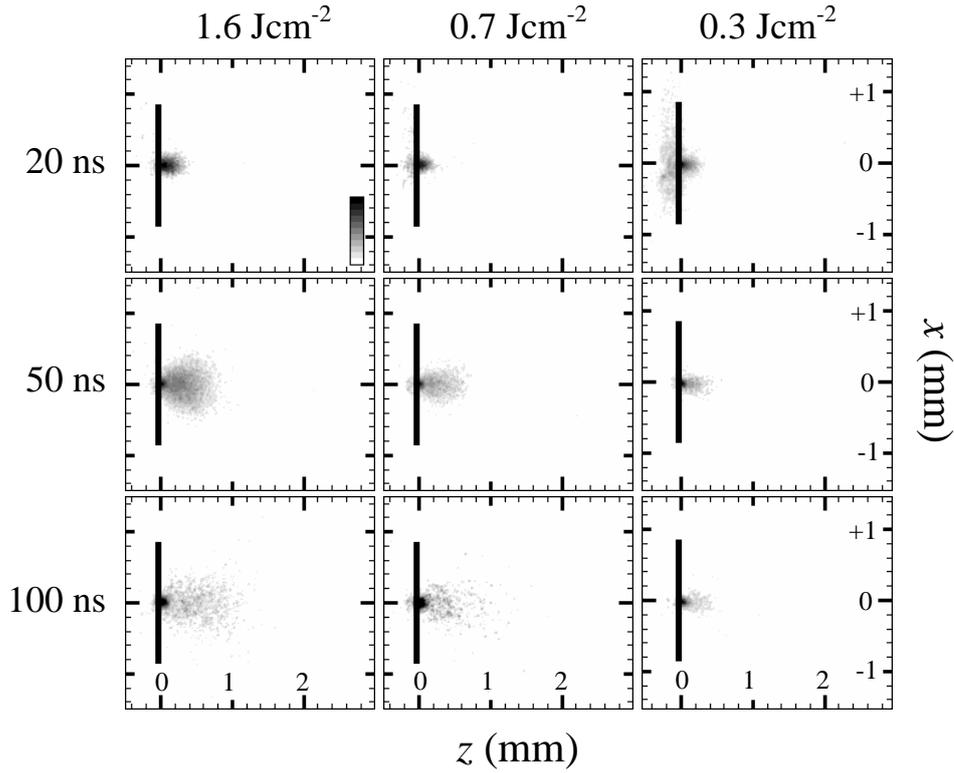

Figure 2 : Plasma images recorded for different laser fluences and time delays during Ti ablation in 500 Pa He using the second harmonic ($\lambda$ = 400 nm). Colour scaling is adjusted for every image to the maximum intensity.

Plasma images recorded during ablation of Ti in 500 Pa He produced by $\lambda_{las}$ = 400 nm (second harmonic) laser pulses are presented in figure 2 for different laser fluences $F_{las}$ and different time delays $t$ between the laser pulse and the observation gate. The intensity scaling (see caption for $F_{las}$ = 1.6 Jcm$^{-2}$, $t$ = 20 ns) has been adjusted to the maximum intensity value for every image. It is noted that the plasmas emission intensity is a strongly decreasing function of time (over several orders of magnitude during 1 µs). The target position is indicated on each image by a vertical black bar. At $t$ = 20 ns, the plasma is located in a volume of about 0.2 mm diameter in the vicinity of the irradiated area of the targets surface for all laser fluences. Later, at 50 ns, two regions of plasma emission are distinguished. Very close to the targets surface, a small volume ($\leq$ 0.1 mm) of high emission intensity is observed whereas a much larger zone of lower emission intensity and conical shape is detected for larger distance from target. This zone seems to be smaller at lower laser fluence. However, it is noted that the size on the image is correlated to the relative intensity with respect to the maximum intensity (strong emitting zone close to the target). Therefore, care has to be taken when interpreting the plasma images. Finally, at 100 ns, the intensity ratio between the slow plasma component (close to the target) and the fast plasma component further increased in favour of the slower component emission intensity.

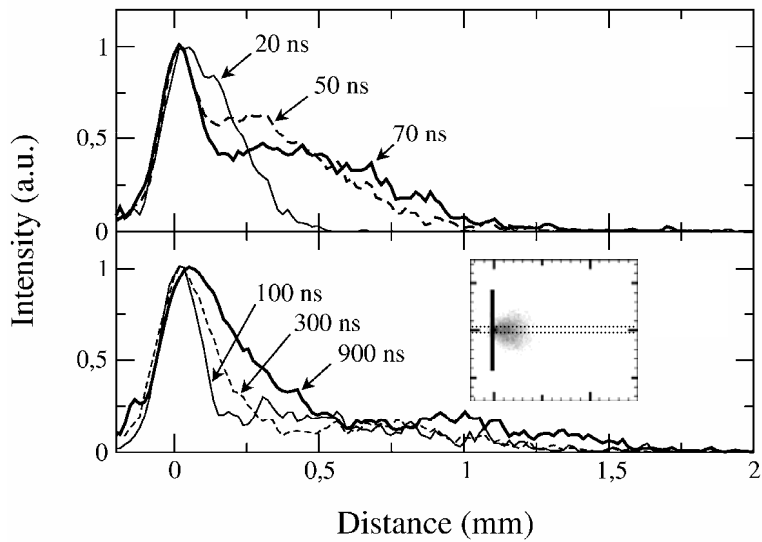

Figure 3 : Intensity as a function distance from target for different times and $F_{las}$ = 1.6 Jcm$^{-2}$. The other experimental conditions are identical to those of figure 2.

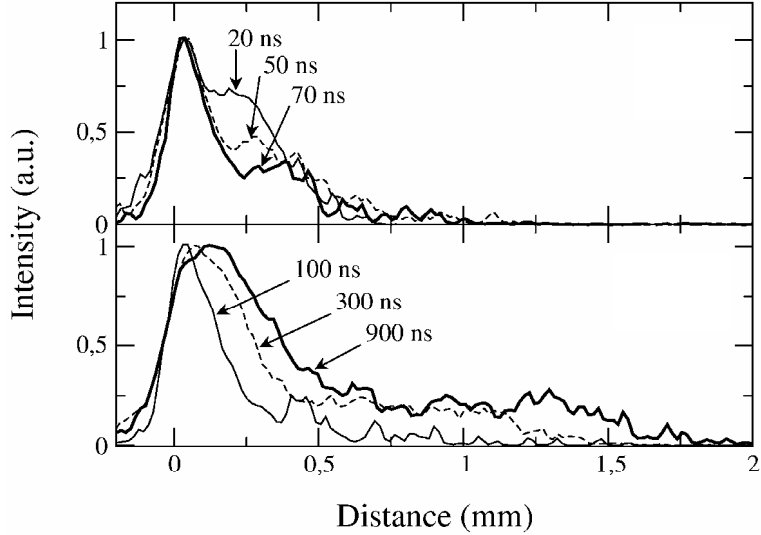

Figure 4 : Intensity profiles recorded during ablation of Ti in 500 Pa He with $F_{las} = 3$ Jcm$^{-2}$ using the fundamental wavelength at $\lambda = 800$ nm.

More detailed information about the two plasma components is obtained from the intensity distribution as a function of distance from target $z$ which is presented in figure 3 for experimental conditions identical to those of figure 2 (1.6 Jcm$^{-2}$). The intensity profiles $I = f(z)$ are obtained from the images by averaging over several lines as shown in the insight of figure 3. It is shown that the intensity profile for $t = 20$ ns already allows one to distinguish between the two plasma components. However, the spatial separation between both components is weak. Therefore, an identification of both components on the image was not possible for $t = 20$ ns. With increasing time, both components show a different expansion behaviour with respect to the $z$-axis. The fast component has an intensity maximum that propagates in time towards higher distances. Contrarily, the slow component is characterised by an intensity maximum that remains close to the targets surface whereas the intensity distribution expands towards higher distances. The most probable velocity of the slow plasma component is about one order of magnitude lower than that of the fast one.

To make sure that the observation of the two plasma components are not an artefact of our experiment but a consequence of the ultra-short laser ablation process itself, we have performed a series of complementary measurements. First, the influence of the applied number of laser pulses on the same surface area has been investigated. As indicated in section 2, the plasma images have been recorded by averaging over several laser ablation events. This has been done by applying a laser pulse sequence onto target which was translated continuously with a velocity of 30 µm s$^{-1}$. Thus, according to the laser spot diameter of 30 µm s$^{-1}$, every surface area was submitted to irradiation by 10$^3$ laser pulses. For a typical ablation rate of 10 nm pulse$^{-1}$, an ablation depth of 10 µm was obtained. In a previous study [6], it was shown that the ablation rate as a function of laser pulse number was constant for ablation depths smaller than the laser spot diameter. Thus, no change of the ablation regime due to the application of successive laser pulses was expected. To make sure that the properties of the plasma plume produced by multipulse ablation are identical to a single shot (fresh surface) generated plume, plasma images have been recorded during a single shot ablation event on a fresh surface for several delay times. Comparing these images to those obtained by multipulse ablation, no significant difference has been observed.

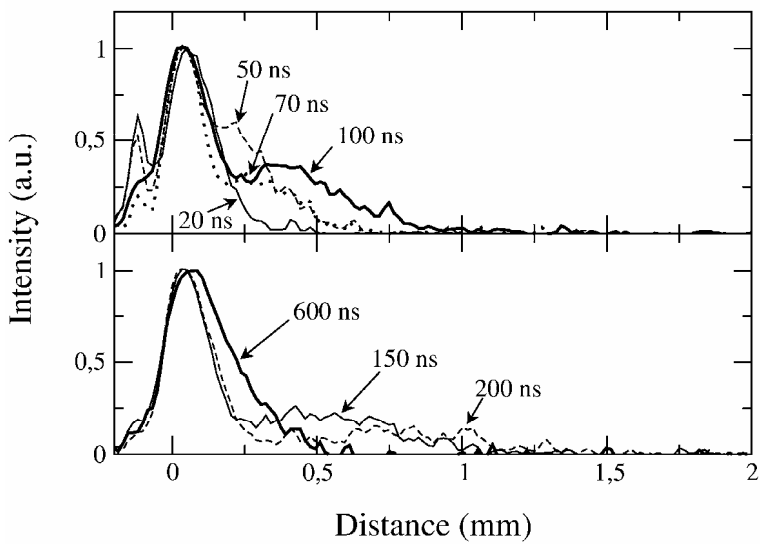

Figure 5 : Intensity profiles recorded for ablation of Ti in vacuum. The other experimental conditions are identical to those of figure 3.

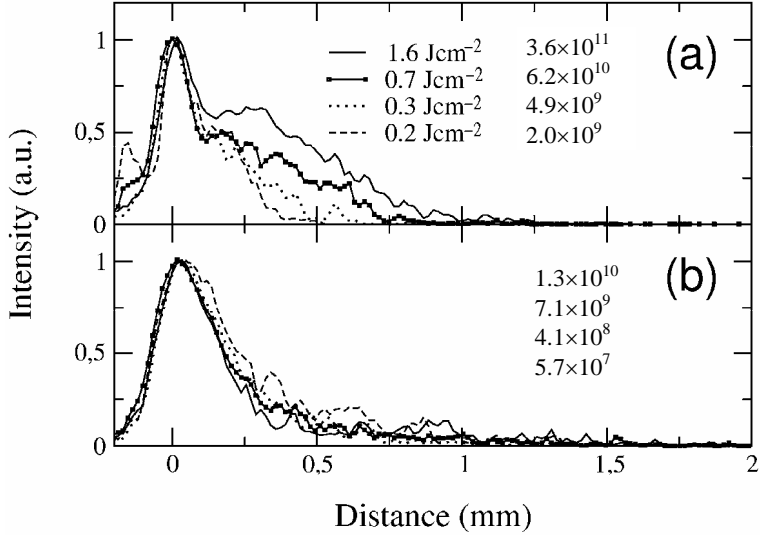

Figure 6 : Intensity profiles recorded for different laser fluences for $t = 20$ (a) and 400 ns (b) during ablation of Ti in 500 Pa He with $\lambda_{las} = 400$ nm.

Second, the influence of the laser pulse properties has been examined. In fact, ultra-fast laser systems do not generate a single perfect laser pulse of 100 fs pulse duration. The laser pulse has not a gaussian temporal shape. It is characterised by a fast intensity increase and an almost exponential intensity decay. That means that the ultra-short pulse is accompanied by radiation on a picosecond or nanosecond time scale of much lower intensity. Furthermore, laser systems with a regenerative amplifier generate a pre-pulse which precedes the main pulse by one or several nanoseconds. The intensity ratio between main- and pre-pulse (pre-pulse contrast) is about $10^3$ for the laser system used in our experiment. In order to avoid problems related to the intensity contrast, we carried out most of the experiments by using the second harmonic of the laser beam at $\lambda_{las} = 400$ nm. The non-linear frequency conversion strongly increases the contrast. In figure 4 are presented intensity profiles for different time delay which have been recorded for similar experimental conditions as those presented in figure 3. Dislike to figure 3 (ablation with $\lambda = 400$ nm, 1.6 Jcm$^{-2}$), figure 4 presents results obtained with laser pulses at the fundamental wavelength at $\lambda = 800$ nm and a fluence of 3.0 Jcm$^{-2}$. Comparing figures 3 and 4, we observe a similar behaviour for ablation with both the fundamental wavelength and the second harmonic. In particular, at $\lambda = 800$ nm (figure 4) the slow and the fast plasma components are identified. Both have similar expansion and propagation velocities as those observed at $\lambda = 400$ nm (figure 3). As the contrast at the fundamental laser wavelength is much lower than that of the second harmonic, it is concluded that the observed plasma propagation behaviour is not an artefact resulting from a low contrast of the used laser source.

The third possible origin of a two component formation in the ablation plume may come from the interaction with the surrounding low pressure gas. Although the mean free path of the ablated species in He at a pressure of 500 Pa is about 100 µm and the effect of the gas on the plume is expected to be small during the early expansion stage, we performed complementary measurements in vacuum to exclude any doubt on the origin of the observed plasma behaviour. In figure 5 are presented intensity profiles for different time delay which have been recorded for experimental conditions identical to those of figure 3, except of the ambient gas pressure which was reduced to $10^{-4}$ Pa. It is seen that both slow and fast plasma components are still observed in vacuum and are thus not produced by the interaction of the ablation plume with the low pressure gas atmosphere. It is however noted, that in vacuum, the emission intensity of the fast component strongly decreases for $t > 200$ ns. This is attributed to the fact, that the density of the expanding plasma plume density strongly diminishes as a function of time. Thus, the collision frequencies decrease and the excitation rates are small during the later expansion stage in vacuum. We observed, however, that the excitation rates in presence of an ambient gas are increased due to collisions between ablated vapour species and ambient gas atoms.

Figure 6 shows intensity profiles for different laser fluences and two different delay times. The times $t = 20$ (a) and 400 ns (b) are chosen to show the influence of laser fluence on the fast and the slow plasma component, respectively. All intensity profiles are normalised with respect to its maximum intensity to allow an easy comparison of the curve shapes. The maximum intensities are given in the figure in counts s$^{-1}$. It is observed that for $F_{las} \geq 0.7$ Jcm$^{-2}$ the intensity distribution attributed to the fast component (a) has a similar shape which indicates that the corresponding propagation velocity is independent of $F_{las}$ in this fluence range. However, for $F_{las} < 0.7$ Jcm$^{-2}$, the intensity distribution becomes more narrow showing that the average propagation velocity of the fast component is diminished. On the contrary, the shape of the intensity profile attributed to the slow component does not depend on laser fluence as shown in figure 6 (b). Thus, the slow component's velocity distribution is not affected by the laser fluence, and only the emission intensity changes. The rise of plasma emission intensity with laser fluence may have different origins. First, the number of emitting species growths up due to the increase of ablation rate. Second, the plasma temperature may increase with laser fluence, which also leads to higher excitation rates. The latter effect is particularly important for laser ablation with nanosecond laser pulses where the laser pulse interacts with the plasma plume and heats up the ablated material. Assuming a linear dependence between laser fluence and ablation rate, the number of emitters increases linearly with $F_{las}$. For constant temperature, an intensity increase $I \propto F_{las}^{\beta}$ would be expected with $\beta = 1$ or $\beta < 1$ for the case of an optically thin or optically thick plasma, respectively. However, the intensity increase observed in figure 6 (a) for $t = 20$ ns is characterised by $\beta > 1$ and thus is stronger than the expected variation for constant $T$. The temperature thus is supposed to increase with laser fluence.

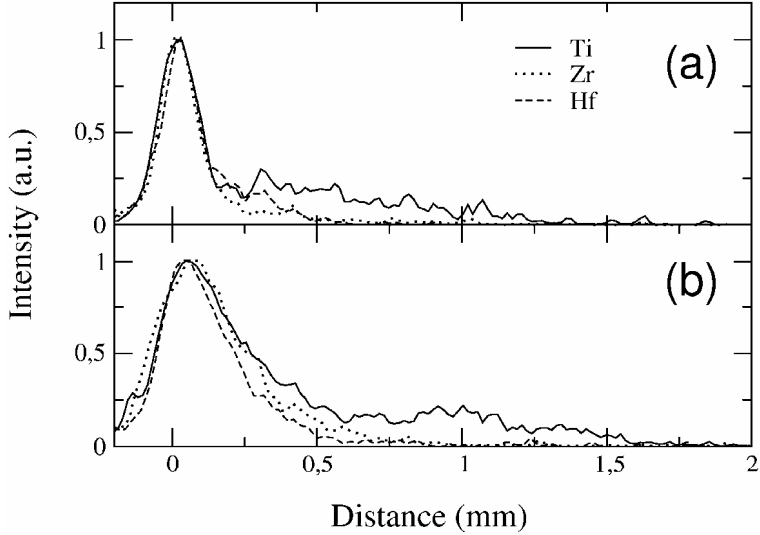

Figure 7 : Intensity profiles recorded during ablation of Ti, Zr and Hf for $t = 100$ (a) and 900 ns (b) in 500 Pa He with $F_{las} = 1.6$ Jcm$^{-2}$ and $\lambda_{las} = 400$ nm.

Intensity profiles of plasma emission recorded during ablation of Ti, Zr and Hf are presented in figure 7 for $t = 100$ (a) and 500 ns (b). It is noted that with respect to the emission intensity of the fast component of Ti, those of Zr and Hf are much weaker. This indicates, that the proportion of fast species with respect to slow ones is reduced for the heavier metals. Furthermore, the more narrow intensity shapes of the fast components of Zr and Hf show that the attributed average velocities are reduced with respect to the lighter Ti (see figure 7 (a)). The expansion velocity of the slow component depends only very weakly on the atomic mass of the ablated material (see figure 7 (b)). For a thermal ablation mechanism, the expansion velocity depends on atomic mass as $u \propto M^{-1/2}$. The observed weaker dependence may be explained by an increase of plasma temperature with the atomic mass. This is a reasonable assumption, because for the Ti, Zr and Hf, the evaporation temperature increases with $M$.

### 3.2. Optical emission spectroscopy

Typical emission spectra of the plasma generated by laser ablation of Zr are presented in figure 8. The spectra have been recorded at a distance of 150 µm from target for $t = 30$ (a) and 70 ns (b). It is observed that the relative intensity of spectral line emission of ions with respect to that of neutral atoms is higher at early times. This shows that the ionisation temperature is higher for the faster plasma species. As a general result, only spectral lines of atoms and single-charged ions were detected during ablation of the three metals. Although particular attention was paid to the research of double-charged ions, a complete absence of emission from these species was noted even for the highest laser fluence of 3 Jcm$^{-2}$. The spectral lines of atoms and ions, which have been selected for time-of-flight analyses, are listed in table 2.

| Species | $\lambda$ (nm) | $A_{ul}$ (10$^8$ s$^{-1}$) | $E_l$ (cm$^{-1}$) | $J_l$ | $E_u$ (cm$^{-1}$) | $J_u$ (cm$^{-1}$) |
|---|---|---|---|---|---|---|
| Ti I | 367.17 | 0.0459 | 386.8 | 4.0 | 27614.6 | 4.0 |
| Ti I | 399.86 | 0.407 | 386.8 | 4.0 | 25388.3 | 4.0 |
| Ti I | 498.17 | 0.659 | 6842.9 | 5.0 | 26910.7 | 6.0 |
| Ti II | 368.52 | 0.746 | 4897.6 | 3.5 | 32025.4 | 2.5 |
| Zr I | 354.76 | 0.536 | 570.4 | 3.0 | 28749.8 | 4.0 |
| Zr II | 357.30 | 0.0791 | 2572.2 | 1.5 | 30551.4 | 2.5 |
| Hf I | 417.43 | 0.0672 | 2356.6 | 3.0 | 26305.7 | 3.0 |
| Hf II | 356.17 | 1.09 | 0.0 | 1.5 | 28068.7 | 1.5 |

Table 2 : Spectroscopic constants of the observed transitions.

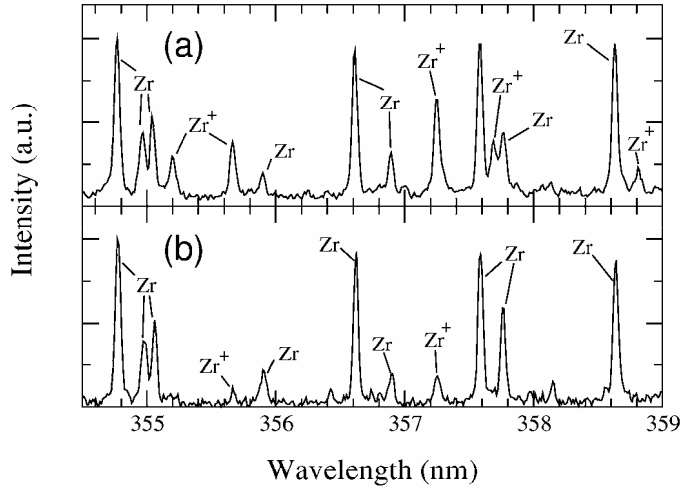

Figure 8 : Optical emission spectra recorded during ablation of Zr in 500 Pa He with $F_{las} = 3.0$ Jcm$^{-2}$ ($\lambda_{las} = 800$ nm) for $t = 50$ (a) and 70 ns (b) and $z = 150$ µm.

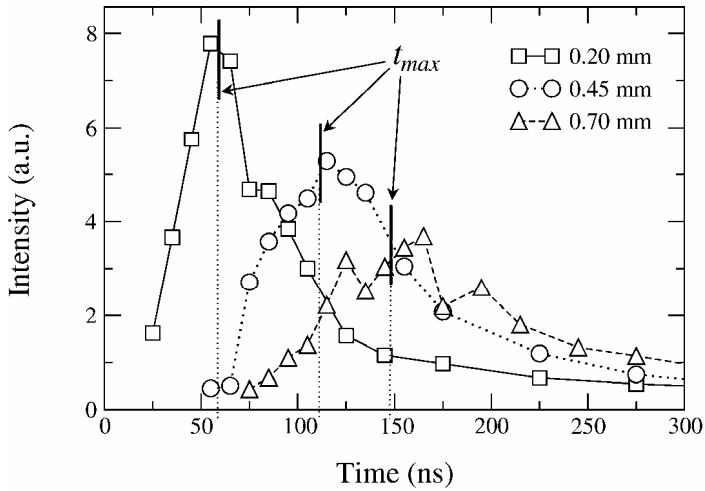

Figure 9 : Emission intensity of the Hf I 417.43 nm spectral line as a function of time for different distances from target. The recordings correspond to ablation of Hf in 500 Pa He with $F_{las} = 3.0$ Jcm$^{-2}$ ($\lambda_{las} = 800$ nm).

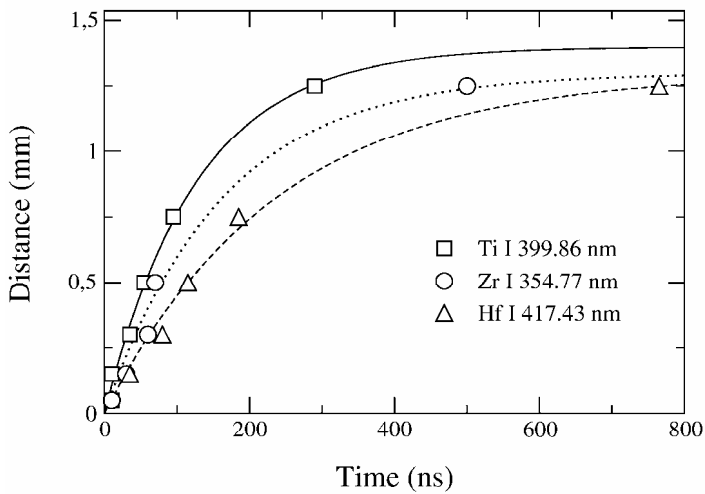

Figure 10 : Distance from target as a function of $t_{max}$ for ablation of Ti, Zr and Hf in 500 Pa He with $F_{las} = 3.0$ Jcm$^{-2}$ ($\lambda_{las} = 800$ nm).

An example of the time-of-flight measurement is given in figure 9, where the emission intensity of the Hf I 417.43 nm spectral line is presented as a function of time for different distances from target. As the emission intensity strongly decreases with $z$, the $z = 0.45$ and 0.70 mm curves have been multiplied by a factor of 2 and 3, respectively, to facilitate the comparison of temporal shapes. It is shown that the time $t_{max}$ (see vertical bars), when the emission intensity reaches its maximum value, increases with distance from target. The time $t_{max}$ has been measured for several distances for the spectral lines listed in table 1. The observation distance is presented as a function of $t_{max}$ in figure 10 for spectral lines of neutral atoms. The slope of the $z = f(t_{max})$ represents the most probable velocity $u_p = \Delta z / \Delta t_{max}$ [7]. The continuous curves in figure 10 have been obtained from calculations using the Drag-force model which has been previously employed to describe the interaction of an expanding vapour plume produced by nanosecond laser ablation with a low-pressure ambient gas [8]. The model is based on the assumption that the viscous force is proportional to the propagation velocity of the ablated species. It leads to an expansion behaviour which is described by

$$z = z_0 \left(1 - e^{-u_0 t / z_0}\right) \qquad (1)$$

where $u_0$ is the initial propagation velocity and $z_0$ the so-called stopping distance. It is shown in figure 10 that the stopping distance is larger than 1 mm and that the expansion velocity is nearly constant for $z < 0.7$ mm. According to this behaviour, the initial expansion velocities have been determined from time-of-flight analyses in the vicinity of the targets surface. The determination of the most probable velocities from the slopes $u_p = \Delta z / \Delta t_{max}$ in the near target region is shown in figure 11 for neutral atoms (a) and ions (b) of the three metals. The values of most probable velocities and the corresponding kinetic energies are listed in table 3. It is emphasised that the propagation velocities of ions and atoms deduced from time-of-flight optical emission spectroscopy correspond to the velocity of the fast plasma component deduced from fast ICCD plasma imaging (see section 3.1). It is shown in table 2 that the propagation velocity varies as $u_p \propto M^{-1/2}$, so that the kinetic energy is independent of the atomic mass.

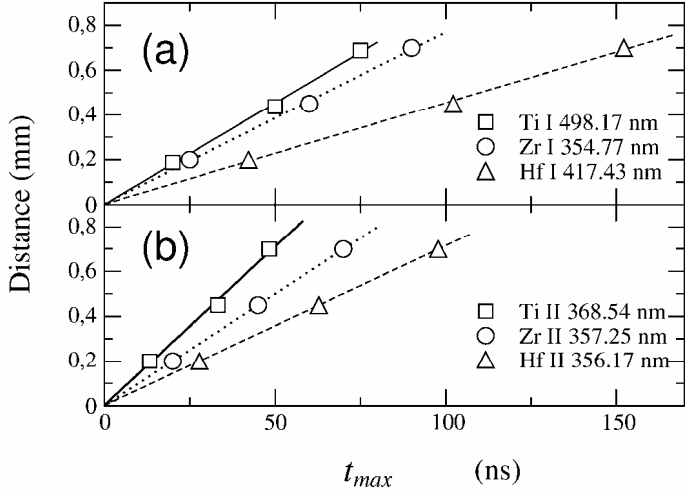

Figure 11 : Distance from target as a function of $t_{max}$ for ablation of Ti, Zr and Hf in 500 Pa He with $F_{las}$ = 3.0 Jcm$^{-2}$ ($\lambda_{las}$ = 800 nm). The slopes represent the most probable velocities of neutral atoms (a) and ions (b).

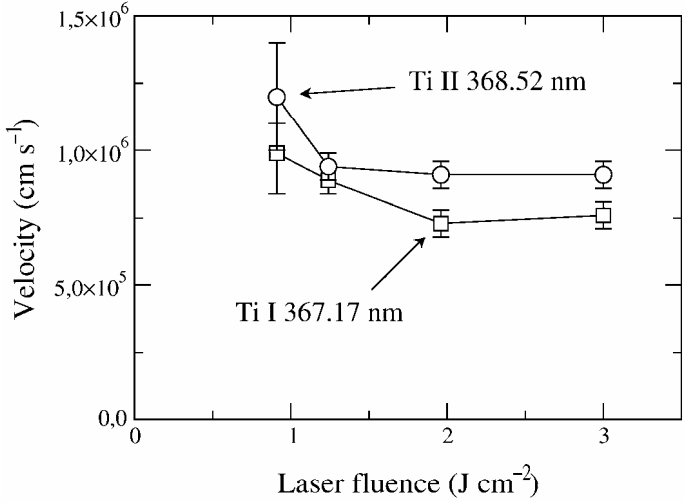

Figure 12 : Propagation velocities of neutral atoms and ions as a function of $F_{las}$ observed during Ti ablation in 500 Pa He using $\lambda_{las}$ = 800 nm.

The influence of laser intensity on the most probale velocity of Ti ions and neutral atoms is presented in figure 12. It is shown that the velocities do not increase with $F_{las}$. It is however noted that this somewhat surprising result has been observed for a fluence range from 0.9 to 3.0 Jcm$^{-2}$. Below 0.9 Jcm$^{-2}$, the plasma emission intensity strongly decreased and the time-of-flight emission spectroscopic measurements were difficult in this fluence range.

The intensity of spectral lines of Ti ions and neutral atoms are presented in figure 13 as a function of $F_{las}$. As a reference value, the ablation threshold fluence of 0.09 Jcm$^{-2}$ is indicated on the laser fluence axis. The measurements correspond to an observation distance of 0.45 mm and to a time delay $t_{max}$ corresponding to the maximum intensity of ionic spectral lines. Thus, the intensity measurements correspond to the highly ionised plasma front which precedes the majority of neutral atoms. As stated before, the emission intensity of both neutral and ionic emission strongly increases with $F_{las}$ in the lower fluence range ($F_{las} \leq 0.9$ Jcm$^{-2}$). For larger fluences, the ratio remains roughly constant. It is stressed that the fluence value of 0.9 Jcm$^{-2}$, corresponding to the change of the intensity increase with $F_{las}$, is about one order of magnitude larger than the ablation threshold. However, $F_{las} = 0.9$ Jcm$^{-2}$ is the value for which a change of ablation regime has been predicted in literature.

From the intensity ratio of ionic and atomic lines, the plasma temperature is estimated to a value of about 5000 K. A laser fluence independent plasma temperature at a distance of about 0.45 mm, which is relatively large with respect to the laser spot diameter of about 30 µm, is not a surprising result. It has been shown by Laville et al. [9], that due to radiative

| Spectral line | $M$ (a.m.u.) | $u_p$ (10$^6$ cm s$^{-1}$) | $E_{kin}$ (eV) |
|---|---|---|---|
| Ti I 498.17 nm | 47.90 | 0.91 | 20.7 |
| Zr I 354.77 nm | 91.20 | 0.75 | 26.8 |
| Hf I 417.43 nm | 178.50 | 0.45 | 18.9 |
| Ti II 368.54 nm | 47.90 | 1.42 | 49.7 |
| Zr II 357.25 nm | 91.20 | 1.00 | 47.7 |
| Hf II 356.17 nm | 178.50 | 0.71 | 47.0 |

Table 3 : Observed transition, atomic mass $M$, most probable velocity $u_p$ and kinetic energy $E_{kin}$ of ablated species.

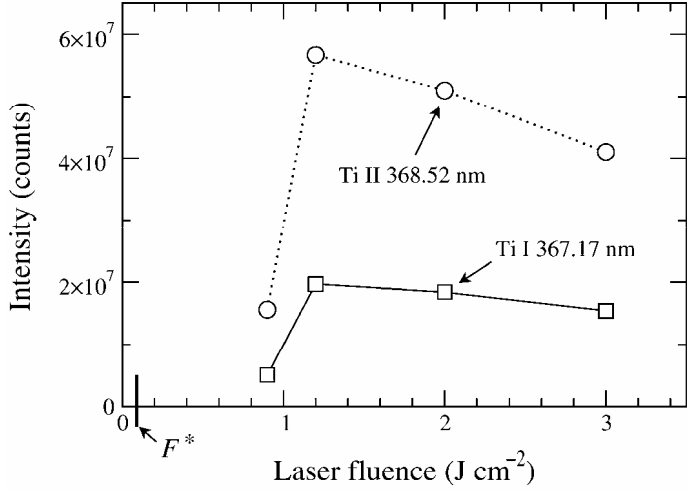

Figure 13 : Spectral line emission intensities of ions and neutral atoms as a function of $F_{las}$ for Ti ablation in 500 Pa He at $\lambda_{las} = 800$ nm.

losses, the temperature evolution of pulsed laser-induced plasmas is characterised by a decrease

$$T(t) = \frac{T_0}{\left(1 + T_0^3 \, A t\right)^{1/3}}, \quad (2)$$

where $T_0$ is the initial temperature while $A$ is numerical constant which depends on the ablated material properties. For large delay times $[t \gg (T_0^3 A)^{-1}]$, the temperature becomes independent of $T_0$. Thus, an eventual increase of the initial temperature with $F_{las}$ does not necessarily increase the temperature at $z = 0.45$ mm measured several tens of nanosecond after the ablating laser pulse. Therefore, time-of-flight measurements of atoms and ions are a more efficient tool to get information on the ablation process itself as the propagation velocities are time-independent due to momentum conservation.

## 4. SUMMARY AND CONCLUSIONS

The present results obtained by fast ICCD plasma imaging and time- and space-resolved optical emission spectroscopy show that the plasma generated by ultra-short laser ablation of Ti, Zr and Hf is characterised by two well-distinguished components. A fast component formed by ions and neutral atoms propagates with a most probable velocity of the order of $10^6$ cm s$^{-1}$ and an opening angle of about 20 to 30°. A slow component has a very wide opening angle ($\geq 80°$) and an average expansion velocity of the order of $10^5$ cm s$^{-1}$. It has been shown that the formation of these two components is neither due the interaction of the ablated material with the low-pressure He atmosphere nor an artefact produced by an eventual too low contrast of the femtosecond laser pulse. The behaviour of both components has been studied as a function laser fluence and target material. Ti, Zr and Hf have similar thermophysical properties but different atomic mass. The propagation velocity was found to vary as $u_p \propto M^{-1/2}$ so that the kinetic energy was independent of $M$. The slow component showed a weak increase of kinetic energy with $M$. The behaviour is an indication of a thermal ablation process because the evaporation temperature - and therefore the expected initial plasma temperature - of the three metals increases with $M$.

The emission spectra of the three metals revealed spectral lines of single-charged ions and neutral atoms. Even for the highest laser fluence of 3 Jcm$^{-2}$, which is about 30 times higher than the ablation threshold, no double-charged ions were detected. Furthermore, the emission intensity of ionic spectral lines and intensity ratio between lines of ions and atoms in the fast propagating plasma front were independent on laser fluence in the range from 1.2 to 3 Jcm$^{-2}$. This behaviour is quite different from that observed during nanosecond laser ablation where the plasma temperature and thus the ionisation degree strongly increases with $F_{las}$ due to plasma heating via inverse Bremsstrahlung [10]. It can be related to energy losses due to fast electron heat diffusion into the material bulk which is supposed to play an important role in the femtosecond ablation process in the fluence regime $F_{las} \geq 1$ Jcm$^{-2}$ [11]. However, a clear identification of both plasma components is not possible at this stage of our study and complementary measurements are planned to get a better understanding of the mechanisms involved into the ultra-fast laser ablation process.

[*]hermann@lp3.univ-mrs.fr ; phone 33 4 91 82 92 90 ; fax 33 4 91 82 92 90 ; www.lp3.univ-mrs.fr